\documentclass{ieeeaccess}
\usepackage{cite}
\usepackage{amsmath,amssymb,amsfonts}
\usepackage{algorithmic}
\usepackage{graphicx}
\usepackage{textcomp}
\usepackage{soul}
\def\BibTeX{{\rm B\kern-.05em{\sc i\kern-.025em b}\kern-.08em
    T\kern-.1667em\lower.7ex\hbox{E}\kern-.125emX}}
\begin{document}
\history{Date of publication xxxx 00, 0000, date of current version xxxx 00, 0000.}
\doi{10.1109/ACCESS.2020.3037415}

\title{Optimized lockdown strategies for curbing the spread of COVID-19: A South African case study}
\author{\uppercase{Laurentz E. Olivier}\authorrefmark{1,3},
\uppercase{Stefan Botha\authorrefmark{2}, and Ian K. Craig}\authorrefmark{3},
\IEEEmembership{Senior Member, IEEE}}
\address[1]{Technology Services, Moyo Africa, 
   Centurion, South Africa.}
\address[2]{Advanced Process Control, Sasol Synfuels, 
   Secunda, South Africa.}
\address[3]{Department of Electrical, Electronic and Computer Engineering, University of Pretoria, 
   Pretoria, 0002, South Africa.}

\markboth
{Author \headeretal: Preparation of Papers for IEEE TRANSACTIONS and JOURNALS}
{Author \headeretal: Preparation of Papers for IEEE TRANSACTIONS and JOURNALS}

\corresp{Corresponding author: Ian K. Craig (e-mail: ian.craig@up.ac.za).}

\begin{abstract}
To curb the spread of COVID-19, many governments around the world have implemented tiered lockdowns with varying 
degrees of stringency. Lockdown levels are typically increased when the disease spreads and reduced when the disease 
abates. A predictive control approach is used to develop optimized lockdown strategies for curbing the spread of 
COVID-19. The strategies are then applied to South African data. The South African case is of interest as 
the South African government has defined five distinct levels of lockdown, which serves as a discrete control input. An epidemiological model for the spread of COVID-19 in 
South Africa was previously developed, and is used in conjunction with a hybrid model predictive controller to 
optimize lockdown management under different policy scenarios. Scenarios considered include how to flatten the curve 
to a level that the healthcare system can cope with, how to balance lives and livelihoods, and what impact the 
compliance of the population to the lockdown measures has on the spread of COVID-19.
The main purpose of this paper is to show what the optimal lockdown level should be given the policy that is in place, as determined by the closed-loop feedback controller.
\end{abstract}


\titlepgskip=-15pt

\maketitle

\hyphenation{mar-gi-nal Bank-serv-Af-ri-ca lock-down}
\bibliographystyle{ieeetran}

\section{Introduction}
\label{sec:introduction}
A novel coronavirus, believed to be of zoonotic origin, emerged in Wuhan, China towards the end of 2019. This virus, which was subsequently named SARS-CoV-2 and the disease it causes COVID-19 \cite{WHO1}, has since spread around the world. The WHO characterized COVID-19 as a pandemic on 11 March 2020 \cite{WHO2}. The COVID-19 pandemic first took hold in regions of the world that share high volumes of air traffic with China \cite{Lau:20}. The importance of ``flattening the curve'', i.e. reducing the number of COVID-19 infected patients needing critical care to be below the number of available beds in intensive care units or appropriately equipped field hospitals, soon became evident \cite{Stewart:20}.

The South African National Institute for Communicable Diseases confirmed the first COVID-19 case in South Africa on 5 March 2020. Having learnt from elsewhere about the importance of ``flattening the curve'', the South African Government was quick to place the country under strict lockdown (what later became known as lockdown level 5) on 27 March 2020 after only 1,170 confirmed COVID-19 cases and 1 related death \cite{HDX:20}.

The early strict lockdown measures in South Africa have been successful from an epidemiological point of view, but great harm was done to an economy that was already weak before the COVID-19 pandemic started \cite{Arndt:20}. As a result, significant pressure was applied to relax the lockdown measures even though the number of infectious individuals was still growing exponentially \cite{BBC}. 

The South African government has formulated five lockdown levels with varying degrees of strictness with regards to the measures imposed in order to systematically restore economic activity. There is thus significant interest in determining the epidemiological impact of the lockdown levels \cite{LockdownLevels}. Towards this end, an epidemiological model was developed for South Africa in \cite{Olivier:20}, and a predictive control approach to managing lockdown levels is presented in this work.

One policy approach to managing lockdown levels may be to flatten the curve so as not to overwhelm the healthcare system. Some infected individuals need hospitalization and intensive care - there are studies that show that roughly 5~\% of confirmed infectious cases require admission to intensive care units (ICUs) \cite{Guan:20}. As more and more individuals are exposed and infected, healthcare systems can easily become overwhelmed, especially in developing countries with fragile and underdeveloped healthcare systems \cite{Ghsindex:20}. The amount of ICU beds available can therefore serve as a high limit for the number of infected individuals needing intensive care \cite{Stewart:20}.

The lockdowns that enforce social distancing however have a huge economic impact \cite{Chamola:20}. Considering the cumulative impact of lockdown on the economy is therefore also relevant. Another policy approach may therefore be to not impose strict lockdown measures for too long, and to potentially reduce the lockdown level even though the healthcare capacity may be exceeded. This policy is known as ``balancing lives and livelihoods'' (see \cite{Panovska:20}).

The aim of this paper is to illustrate the optimal implementation of these policy choices through the use of a model predictive control approach \cite{Camacho:13}. Lockdown levels are represented as integer values, whereas the SEIQRDP model used is continuous and dynamic. Dynamic systems that contain continuous and discrete state/input variables are known as hybrid systems \cite{Camacho:10}, and therefore a hybrid model predictive control (HMPC) approach is required.

Solving the resulting constrained optimization problem is known as mixed integer programming. These problems are NP-hard (non-deterministic polynomial-time) and even to test if a feasible solution improves on the best solution to date is an NP problem \cite{Camacho:10}. Genetic algorithms (see e.g. \cite{Fleming:02}) have been found in the past to be suitable for solving HMPC problems, and are therefore used in this work (see e.g. \cite{Muller:17,Botha:18}).
 
Genetic algorithms (GAs) are founded on the principles of natural selection and population genetics \cite{Fleming:02}. A GA solves the optimization problem in a derivative-free manner using a population of potential solutions that are evolved over generations to produce better solutions. Each individual in the population is assigned a fitness value that determines how well it solves the problem and hence how likely it is to propagate its characteristics to successive populations. A GA does not mathematically guarantee optimality, but provides a feasible solution in an appropriate time frame. Given the limited number of possible control moves, constraint functions, and population size used here, optimality is however likely.

Other approaches to solving HMPC problems are also possible, as illustrated in e.g. \cite{Viljoen:20}.

Once the control move is calculated, some time is required for the country to prepare for the new lockdown level. To achieve this the lockdown level to be implemented is calculated some time in advance. A fixed delay between the control move calculation and implementation is not a typical HMPC requirement. It is however required in this instance for practical implementation of the appropriate lockdown level.

Other approaches to feedback control of COVID-19 in various forms exist in literature. \cite{Tsay:20} and \cite{Bin:20} show how implementation of social distancing measures in an ``on-off'' fashion might be used to stem the spread of the virus. This might be effective in limiting the spread, but an ``on-off'' approach to lockdown is difficult to implement practically and after a number of iterations the population may well not wish to comply any longer. \cite{Perkins:20} shows an optimal control strategy to limit deaths in the United States through implementation of social distancing measures. \cite{Kohler:20} presents a robust MPC approach to calculating optimal social distancing measures for Germany. The degree of social distancing is restricted to only change once a week, but how a real-valued degree of social distancing is practically achieved is less straightforward. \cite{Kohler:20} notes that adaptive feedback strategies are required to reliably contain the COVID-19 outbreak. \cite{Stewart:20} shows an ``on-off'' approach as well as real-valued social distancing measures to control the number of patients requiring intensive care to below the available capacity. All of the controllers referenced in this paragraph also operate under a single policy scenario, as opposed to the framework discussed in this work that deals with multiple scenarios.

In this work, predictive control is used to implement various policy scenarios by varying the lockdown level, where each level has its own well-defined set of regulations. How the control action is practically implemented is therefore well determined. The novelty of the paper is therefore the implementation of an HMPC to control under different policy scenarios within a context where the practical implementation of the control output is well defined.

An overview of the epidemiological model used in this work is given in Section~\ref{sec:model_used} along with the parameters for the South African case. The controller design for different policy scenarios is presented in Section~\ref{sec:controller_design}, the results and discussion in Section~\ref{sec:results}, and finally the conclusion in Section~\ref{sec:conclusion}.

\section{Epidemiological model development}
\label{sec:model_used}

It is noted in \cite{Latif:20} that accurate epidemiological models are indispensable for planning and decision making, such as the feedback control described in this work.

Even though the model development is not the main aim of this paper, a model is required for the implemented HMPC. This section is dedicated to showing the model development, before the main contributions of this paper, i.e. that of controller design and simulation, as are discussed in sections \ref{sec:controller_design} and \ref{sec:results}.

\subsection{SEIQRDP model}
\label{sec:SEIQRDP_model}

\begin{figure}
\begin{center}
\includegraphics[width=8.2cm]{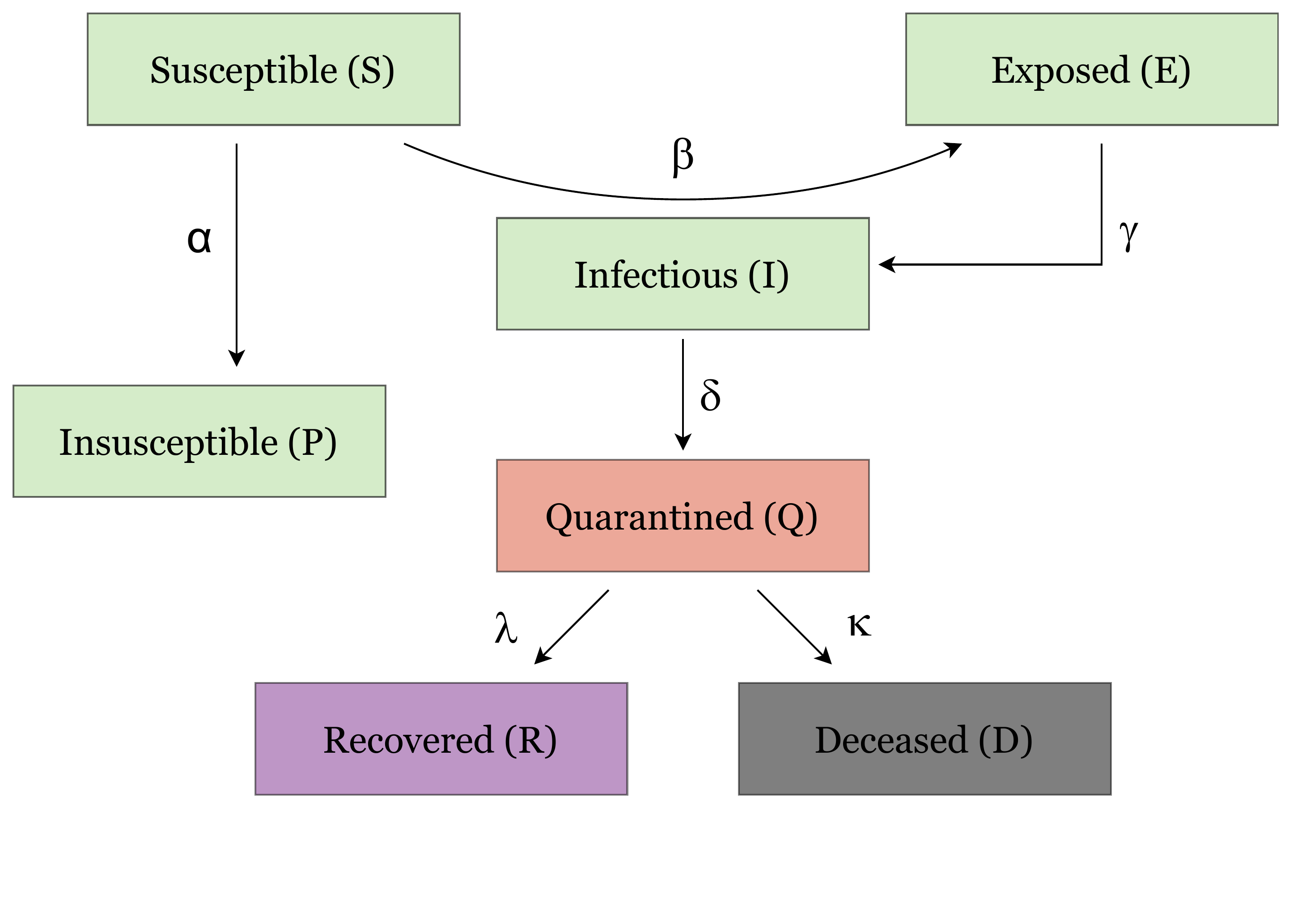}    
\caption{SEIQRDP model proposed by \cite{Peng:20}.} 
\label{fig:SEIQRDP_model}
\end{center}
\end{figure}

The SEIQRDP model is a generalized compartmental epidemiological model with 7 states. The model was proposed by \cite{Peng:20} and is an adaptation of the classical SEIR model (see e.g. \cite{Hethcote:91}). The model states and model parameters that drive transitions between them are shown in Fig.~\ref{fig:SEIQRDP_model}. The colours used for Q, R, and D correspond to what is used in the results figures later in the article. The states are described as:
\begin{itemize}
\item S - Portion of the population still susceptible to getting infected,
\item E - Population exposed to the virus; individuals are infected but not yet infectious,
\item I - Infectious population; infectious but not yet confirmed infected,
\item Q - Population quarantined; confirmed infected,
\item R - Recovered,
\item D - Deceased,
\item P - Insusceptible population.
\end{itemize}

The model equations are given as:

\begin{align}
\label{eq:S}
\frac{dS(t)}{dt} & = -\alpha S(t) - \frac{\beta(t)}{N} S(t)I(t) \\
\frac{dE(t)}{dt} & = -\gamma E(t) + \frac{\beta(t)}{N} S(t)I(t) \\
\frac{dI(t)}{dt} & = \gamma E(t) - \delta I(t) \\
\label{eq:Q}
\frac{dQ(t)}{dt} & = \delta I(t) - \left( \lambda(t) + \kappa(t) \right) Q(t) \\
\frac{dR(t)}{dt} & = \lambda(t) Q(t) \\
\frac{dD(t)}{dt} & = \kappa(t) Q(t) \\
\label{eq:P}
\frac{dP(t)}{dt} & = \alpha S(t)
\end{align}
where $N$ is the total population size, $\alpha$ is the rate at which the population becomes insusceptible (in general either through vaccinations or medication). At the time of writing there is no vaccine that will allow an individual to transfer from the susceptible to insusceptible portion of the population \cite{Promp:20}. Consequently $\alpha$ should be considered to be close to zero. $\beta(t)$ is the (possibly time dependent) transmission rate parameter, $\gamma = [N_{lat}]^{-1}$ is the inverse of the average length of the latency period before a person becomes infectious (in days), $\delta = [N_{inf}]^{-1}$ is the inverse of the number of days that a person stays infectious without yet being diagnosed, $\lambda(t)$ is the recovery rate, and $\kappa(t)$ is the mortality rate. Both $\lambda(t)$ and $\kappa(t)$ are potentially functions of time, and \cite{Peng:20} notes that $\lambda(t)$ gradually increases with time while $\kappa(t)$ decreases with time. As such, the functions shown in (\ref{eq:lambda}) and (\ref{eq:kappa}) are used to model $\lambda(t)$ and $\kappa(t)$. In (\ref{eq:lambda}) it is set that $\lambda_1 \geq \lambda_2$ such that $\lambda(t) \geq 0$.

\begin{align}
\label{eq:lambda}
\lambda (t) & = \lambda_1 - \lambda_2 \exp(-\lambda_3 t) \\
\kappa (t) & = \kappa_1 \exp(-\kappa_2 t).
\label{eq:kappa}
\end{align}

A decreasing $\kappa$-value is observed in the data of various countries (see e.g. \cite{Peng:20}, or the data in \cite{HDX:20} for South Africa or Italy). Even though $\kappa \rightarrow 0$ as $t \rightarrow \infty$, $\kappa > 0$ over the horizon where predictions are made.

$\beta$ is often considered to be constant, but is dependent on interventions like social distancing, restrictions on travel, and other lockdown measures \cite{RSA_Gov:20}. This implies that $\beta$ may also be time dependent as these interventions change over time. In cases where international and even regional travel is limited, the number of susceptible individuals one may encounter can only decrease over time. This, coupled with the fact that given a constant $\delta$, the effective reproduction number (shown in (\ref{eq:R0})) can only reduce to below 1 if $\beta$ decreases. This lead \cite{Olivier:20} to find that $\beta (t)$ can effectively be modelled using a decreasing function of time of the form

\begin{equation}
\label{eq:beta}
\beta (t) = \beta_1 + \beta_2 \exp(-\beta_3 t) . \\
\end{equation}

This function describing $\beta$ might not be applicable for all countries under all circumstances, but for the countries considered in \cite{Olivier:20} it seems reasonable.

The basic reproduction number, $R_0$, which is the expected number of cases directly generated by one case in the population, is given by \cite{Peng:20} as
\begin{equation}
R_0 = \frac{\beta}{\delta} \left( 1 - \alpha \right) ^ T,
\end{equation}
where $T$ is the number of days. When $\alpha \approx 0$, and $\beta$ is a function of time, this can be simplified as
\begin{equation}
\label{eq:R0}
R_0 \approx \frac{\beta (0)}{\delta}.
\end{equation} 

Interventions such as social distancing, restrictions on population movement, and wearing of masks (among others) can reduce the effective reproduction number mainly through reducing the effective number of contacts per person. This is why the imposition of varying lockdown levels may be used as a control handle to effect policy decisions.

Even though compartmental models depend on accurate parameter values that may be time-variant, the model structure is flexible enough to produce relatively accurate descriptions of the spread of the virus for various countries as shown in \cite{Olivier:20}. It is however emphasized that predictions are sensitive to some of the parameter values, and as such need to be taken with caution during the early stages of the epidemic.

\subsection{Parameter estimation for South Africa}
\label{sec:parameter_estimation}

Data are obtained from The Humanitarian Data Exchange\footnote{Accessible from https://data.humdata.org/dataset/novel-coronavirus-2019-ncov-cases}, as compiled by the Johns Hopkins University Center for Systems Science and Engineering (JHU CCSE) from various sources. The data include the number of confirmed infectious cases, recovered cases, and deceased cases per day from January 2020.

In order to get a sense of the applicability of the model and what the parameter values should be, parameter estimations were first carried out to determine SEIQRDP models for Germany, Italy, and South Korea in \cite{Olivier:20}. These countries were selected as their outbreaks started earlier than that of South Africa, and consequently their parameter estimation should be more accurate. They have also had differing approaches under different circumstances, which means that the different parameters obtained should illustrate how the model behaves. Epidemiological considerations largely drove the constraints on parameter values used in the South African model. Parameter values obtained for $\beta$ from the models for other countries however guided the constraints on each parameter in the South African case.

When the number of cases is very low, community transmissions do not drive the spread which makes accurate modelling difficult. As such, the data taken for fitting is from 18 April 2020. The initial ``level 5'' lockdown model derived in \cite{Olivier:20} used data up to 8 May 2020. Here the model is re-fit using data up to 16 August 2020. The South African government relaxed the lockdown to level 4 on 1 May 2020 and to level 3 on 1 June 2020.

To accommodate potentially differing $\beta$-values for different lockdown levels, the effective $\beta$-value is obtained using:

\begin{equation}
\label{eq:beta_per_level}
\beta^{*} (t) = \beta_{m} (l) \cdot \beta (t)
\end{equation}

where $\beta_m(l)$ is the $\beta$-multiplier as a function of the lockdown level ($l$). This achieves a varying contact rate without varying the time dynamics of $\beta(t)$ using a level-based multiplier.

\begin{figure*}
\begin{center}
\includegraphics[width=12cm]{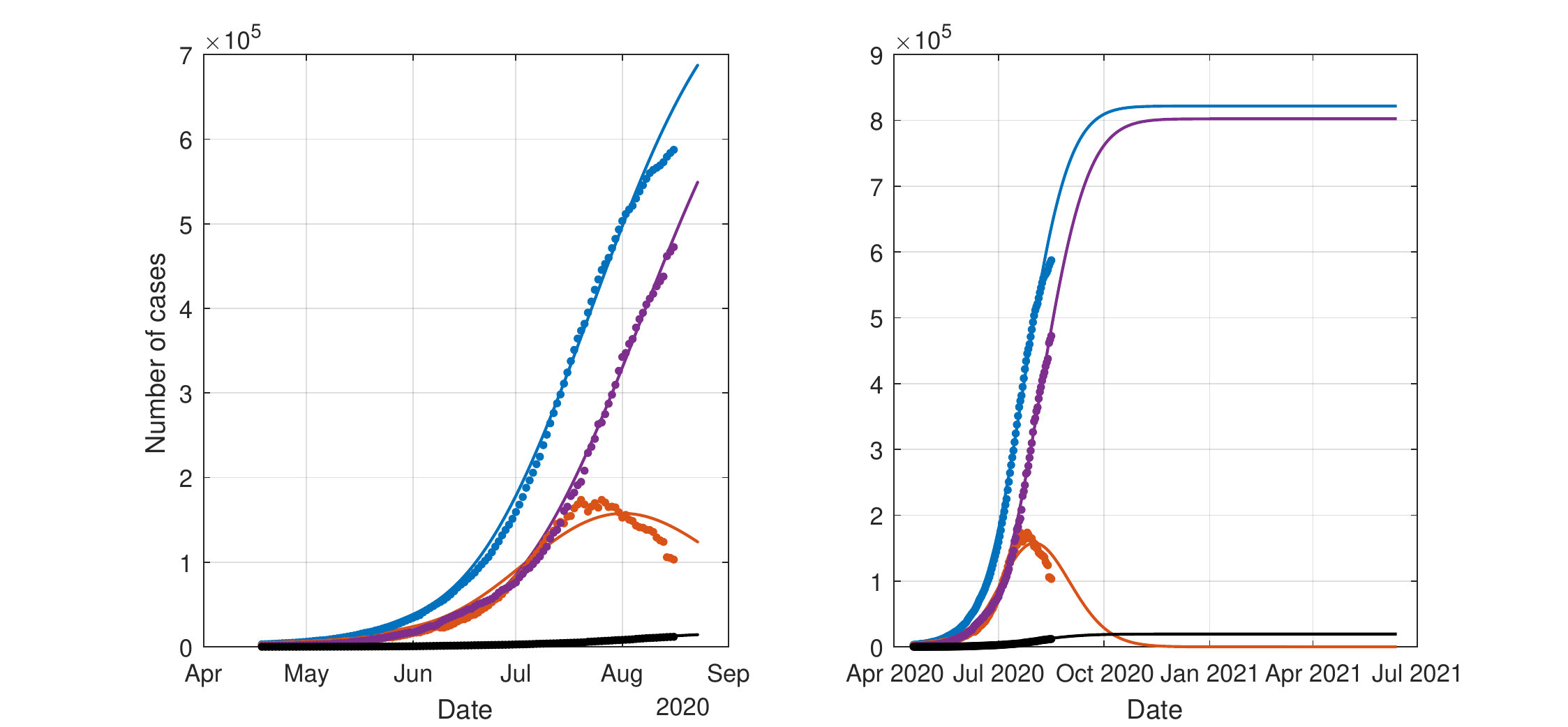}    
\caption{Data fit for South Africa with varying lockdown levels. The figure shows Total cases in blue, Confirmed infectious cases (Q) in orange, Recovered cases (R) in purple, and Deceased cases (D) in black.} 
\label{fig:model_fit_different_levels}
\end{center}
\end{figure*}

Modelling can therefore estimate the $\beta$-values for level 5 ($\beta_m(5)$), level 4 ($\beta_m(4)$), and level 3 ($\beta_m(3)$). The fit of the model to the data is shown in Fig.~\ref{fig:model_fit_different_levels} and the estimated parameter values are shown in Table~\ref{tb:fitting_parameters}.

\begin{table}[t]
\begin{center}
\caption{Model parameters for South Africa using data from 18 April to 16 August 2020.}
\label{tb:fitting_parameters}
\begin{tabular}{ccccc}
Param. & Min & Max &  South Africa \\\hline \\[-7pt]
$\alpha$    & 0     & $\mathrm{10^{-6}}$ & $\mathrm{1 \times 10^{-6}}$ \\
$\beta_1$   & 0     & $R_0 < 3.58$    & 0.011 \\
$\beta_2$   & 0     & $R_0 < 3.58$    & 0.606 \\
$\beta_3$   & 0     & $R_0 < 3.58$    & 0.014 \\ 
$\gamma$    & 0.2   & 1       		  & 0.200 \\
$\delta$    & 0.1   & 1       		  & 0.181 \\
$\lambda_1$ & 0     & 1       		  & 0.113 \\
$\lambda_2$ & 0     & 1       		  & 0.104 \\
$\lambda_3$ & 0     & 1       		  & 0.007 \\
$\kappa_1$  & 0     & 1       		  & 0.002 \\
$\kappa_2$  & 0     & 1       		  & 0.000 \\
$\mathrm{N}$ (million)	& - & - 	  & 57.78 \\[5pt]
\hline \\[-7pt]
MSE & 0 & $\infty$ & $\mathrm{2.09 \times 10^{8}}$ \\
$\mathrm{R^2}$ & $-\infty$ & 1 & 0.991 \\
\hline
\end{tabular}
\end{center}
\end{table}

\begin{figure}
\begin{center}
\includegraphics[width=8.4cm]{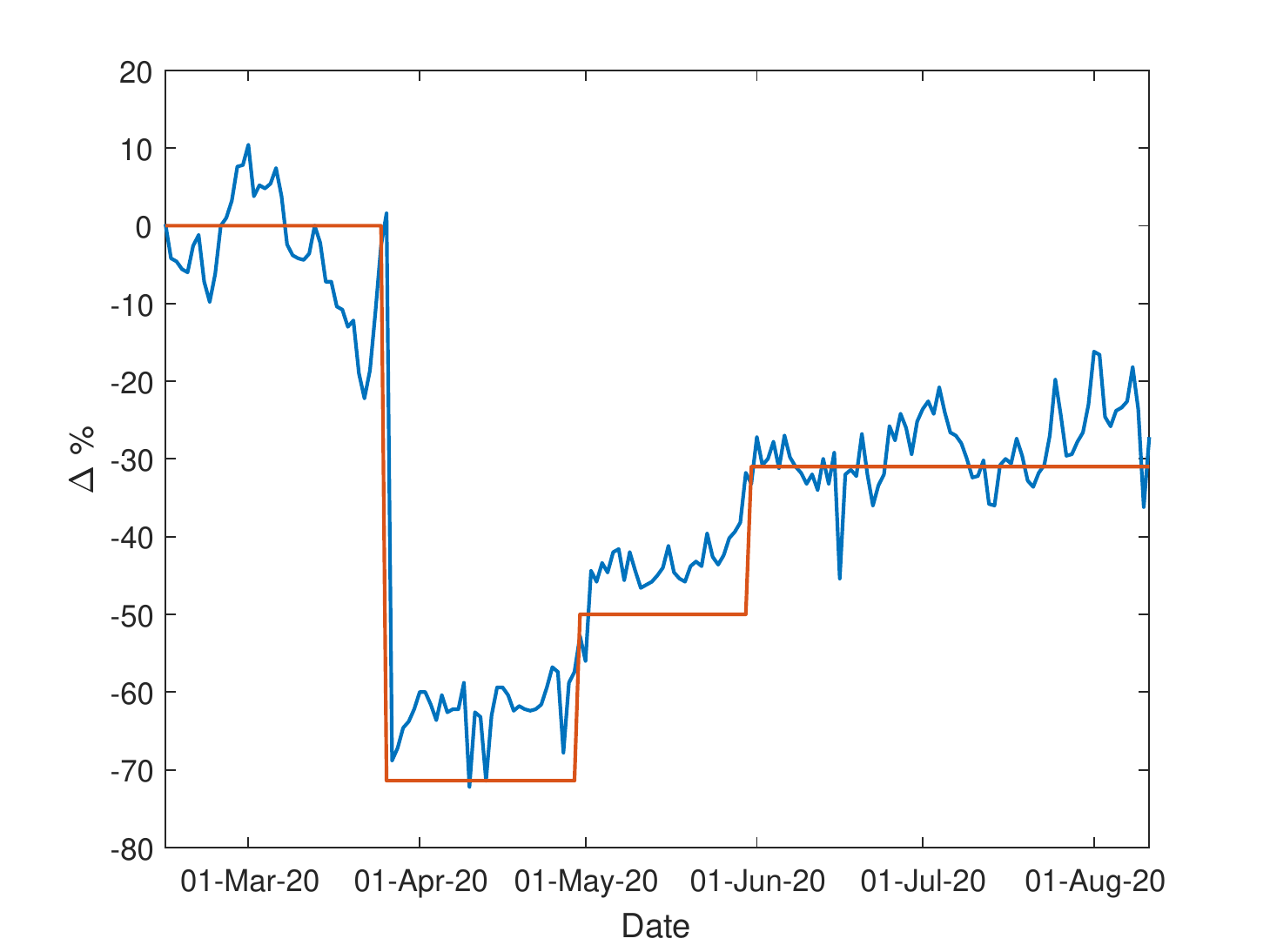}    
\caption{Percentage change in time spent at non-residential locations in blue and percentage change in $\beta$-multiplier per lockdown level in orange.} 
\label{fig:mobility}
\end{center}
\end{figure}

$\beta_m(1)$ and $\beta_m(2)$ still need to be determined. To get a sense of how the contact rate may have changed during different lockdown levels, the Google mobility data for South Africa (obtained from \cite{Marivate:20}) are considered. The mobility data list the relative percentage change in time spent at different location types, based on the median value for the period between 3 January and 6 February 2020. The average difference for all non-residential location types is used to determine an average overall indication of much time people are spending outside of their homes, and in potential contact with others. The index is shown in Fig.~\ref{fig:mobility}. If $\beta_m(1)$ is considered to correspond to the period before level 5 lockdown is implemented, then $\beta_m(2)$ can be calculated such that it lies partway between $\beta_m(1)$ and $\beta_m(3)$.

The values for $\beta_m(l)$ are calculated as such and the final $\beta$-multiplier values are shown in Table~\ref{tb:beta_by_level}. The percentage change in $\beta$-multiplier value from level 1 is also shown in Fig.~\ref{fig:mobility}. It can be seen from Fig.~\ref{fig:mobility} that the $\beta$-multiplier values obtained seem reasonable when compared to mobility data, and that compliance to lockdown measures does decrease over time as the mobility tends to increase while the lockdown level remains constant.

\begin{table}[t]
\begin{center}
\caption{$\beta$-multiplier values per lockdown level.}
\label{tb:beta_by_level}
\begin{tabular}{ccc}
Lockdown level & $\beta_m$ & $\Delta$ from level 1 (\%) \\\hline
1 & 1.638 & -- \\
2 & 1.319 & -19.45 \% \\
3 & 1.000 & -38.96 \% \\
4 & 0.681 & -58.44 \% \\
5 & 0.468 & -71.41 \% \\
\hline
\end{tabular}
\end{center}
\end{table}

One criticism of deterministic epidemiological models (see e.g. \cite{Britton:10}) is that if $R_0 < 1$ there will only be a small outbreak, and if $R_0 > 1$ there will be a major outbreak. This is because the model assumes that the community is homogeneous and that individuals mix uniformly. In reality however individuals will not mix uniformly, especially if regional travel is prohibited. This means that the effective reproduction number does decrease over time in practice, as modelled in \cite{Olivier:20}. Varying $\beta$ however leads to large model sensitivity, meaning that the total number of cases predicted may be higher than what occurs in practice. The varying $\beta$ version of the model is however still useful to illustrate the effect of policy decisions on the relative number of cases recorded.

\section{Controller design}
\label{sec:controller_design}

An HMPC controller using a GA to solve for the optimal control action (similar to \cite{Botha:18}) is implemented to determine optimal lockdown levels for different policy scenarios. The controlled variable is the active number of confirmed infectious cases: $Q$ as given by (\ref{eq:Q}).

The control problem is formulated as:

\begin{equation}
\begin{array}{cl}
\underset{u_k \ldots u_{k+N_c-1}}{\min} & J(u_k, \ldots, u_{k+N_c-1},\mathbf{x}_k) \\
s.t. \; \, & \mathbf{x}_{k+1} = \mathbf{f} \left( \mathbf{x}_{k},u_{k},\mathbf{\theta}_{k} , d_k \right) \\
 & y_k = \mathbf{g} \left( \mathbf{x}_k, \mathbf{\theta}_{k} \right) \\
 & \theta_{c}(y_k \ldots y_{k+N_p},u_k \ldots u_{k+N_c-1})\leq0 
\end{array}
\end{equation}
where $x:\mathbb{R}\rightarrow\mathbb{R}^{n_{x}}$ is the state trajectory, $u:\mathbb{R}\rightarrow\mathbb{R}^{n_{u}}$ is the control trajectory, $\mathbf{x}_{k} = \left[ S_k, E_k, I_k, Q_k, R_k, D_k, P_k \right]$ is the state at time step $k$, $\mathbf{\theta}_{c}(\cdot)$ is a possibly nonlinear constraint function (which is shown in (\ref{eq:nonlcon})), $\mathbf{f}(\cdot)$ is  the state transitions as given in (\ref{eq:S}) - (\ref{eq:P}), $\mathbf{g}(\cdot) = \mathbf{I}_7 \cdot \mathbf{x}$ is the output function and since only Q is controlled may be simplified to be $y_k = Q_k$, $u_k$ contains the exogenous input (the lockdown level \textit{l} as in (\ref{eq:beta_per_level})), $\mathbf{\theta}_k$ represents the system parameters, and $\mathbf{d}_k \in D$ represents the disturbance, which is later introduced in (\ref{eq:compliance_beta}). The performance index (or objective function) to be minimized, $J(\cdot)$, depends on the policy in place as presented in the rest of this section, and is given in (\ref{eq:obj_function_herd_immunity}) and (\ref{eq:obj_function_economic_impact}). Full state feedback is assumed.

\subsection{Flattening the curve policy}
\label{sec:herd_immunity}

A flattening the curve policy is one where lockdown is implemented in order to ensure that the healthcare system is not overwhelmed by keeping the maximum number of cases requiring intensive care below the number of ICU beds available \cite{Kissler:20}. In South Africa the number of ICU beds available for COVID-19 patients was stated in \cite{Cowan:20b} to be 7,188. Using the number of beds for the country as a whole may be somewhat crude. Regional values might be preferred, but as the model is for the country as a whole the high limit is considered in that fashion as well.

\cite{Guan:20} found that roughly 5 \% of confirmed infectious cases required admission to ICU. With 7,188 ICU beds available for COVID-19 patients, this implies that the active confirmed infectious case number should remain below 143,760.

The objective function used to implement this policy penalizes an output (confirmed infectious cases Q) above the number of ICU bed imposed high limit as well as the magnitude of the control move (the lockdown level). This ensures that the output value will tend to remain below the high limit without setting the lockdown level needlessly high. The objective function is stated as:

\begin{equation}
\label{eq:obj_function_herd_immunity}
J(\cdot) =  \displaystyle\sum\limits_{i=1}^{N_p} \| s_i \|_{W_s}^2 + \displaystyle\sum\limits_{i=0}^{N_c-1} \| u_{i} \|_{W_u}^2
\end{equation}
where $N_p$ and $N_c$ are the prediction and control horizons respectively; $\| \cdot \|_W^2$ is the $W$-weighted 2-norm; $W_s$ and $W_u$ are weights corresponding to the relative importance of penalizing slack variables for constraint violations and control values respectively. The slack variables are represented by $s_i$ and are defined to be:
\begin{equation} 
\label{eq:slack_vars}
 s_i = \left\{
\begin{array}{lcl}
 y_i - y_h & ; & y_i > y_h \\
 0 & ; & y_i \leq y_h
\end{array} 
\right.
\end{equation}
where $y_h$ is the output high limit.

The sampling period $\Delta T = 1$ day (which was found to produce a sufficient resolution for progressing the simulation numerically), $N_p=280$ (which is a long enough horizon to capture the dynamics of the spread of the virus, as shown by the modelling results of \cite{Olivier:20}), and $N_c = 3$ (which is an often used control horizon value that provides a good middle ground between controller aggressiveness and resolution to solve the control problem, see e.g. \cite{Bemporad:10}); $N_c$ is however implemented using a blocking vector \cite{Bemporad:10} of $N_b = [7, 7, 7]$ implying that the lockdown level may only change at most every 7 days (1 week).

\subsection{Balancing lives and livelihoods policy}
\label{sec:economic_impact}

When implementing the ``flattening the curve'' policy the lockdown might end up being extremely long, which has an economic impact in itself. Preventing economic activity and therefore preventing certain people from earning a living will likely increase poverty, which in itself leads to life years lost. As such the cumulative economic impact of lockdown should also be considered, which is done by adapting the objective function to be:
\begin{equation}
\label{eq:obj_function_economic_impact}
J(\cdot) =  \displaystyle\sum\limits_{i=1}^{N_p} \| s_i \|_{W_s}^2 + \displaystyle\sum\limits_{i=0}^{N_c-1} \| u_{i} \|_{W_u}^2 + E_{C,k} + \displaystyle\sum\limits_{i=0}^{N_c-1} \| E_{i} \|_{W_E}^2
\end{equation}
where $E_{C,k}$ is the cumulative economic impact of lockdown levels that have already been implemented, $E_{i}$ is the marginal economic impact of lockdown as implemented over the control horizon (as is ultimately reported in Table~\ref{tb:economic_impact_per_lockdown_level}), and $W_E$ is the weight (scalar) relating to the economic impact. At each simulation step the cumulative economic impact is increased as:

\begin{equation}
\label{eq:cumulative_econ_impact}
E_{C,k+1} = E_{C,k} + W_E \cdot E_{i}.
\end{equation}

For $E_{i}$ the quantified relative economic impact per lockdown level is required. One may wish to consider something like the gross domestic product (GDP) or the value add of the industries that may operate during each lockdown level as in \cite{Arndt:20}. GDP figures are often only reported at least one quarter after the fact which does not help in this case. The percentage of each industry that will be affected per lockdown level is not known directly and estimating these may add too much uncertainty. A more frequently updated indicator, which is used here instead, is the BankservAfrica Economic Transactions Index (BETI) which show the volumes and values of inter-bank transfers \cite{BankServ:20}. BankservAfrica states that BETI is a leading indicator for the South African GDP -- it correlates well with GDP figures while appearing a quarter earlier.

\begin{figure}
\begin{center}
\includegraphics[width=8.4cm]{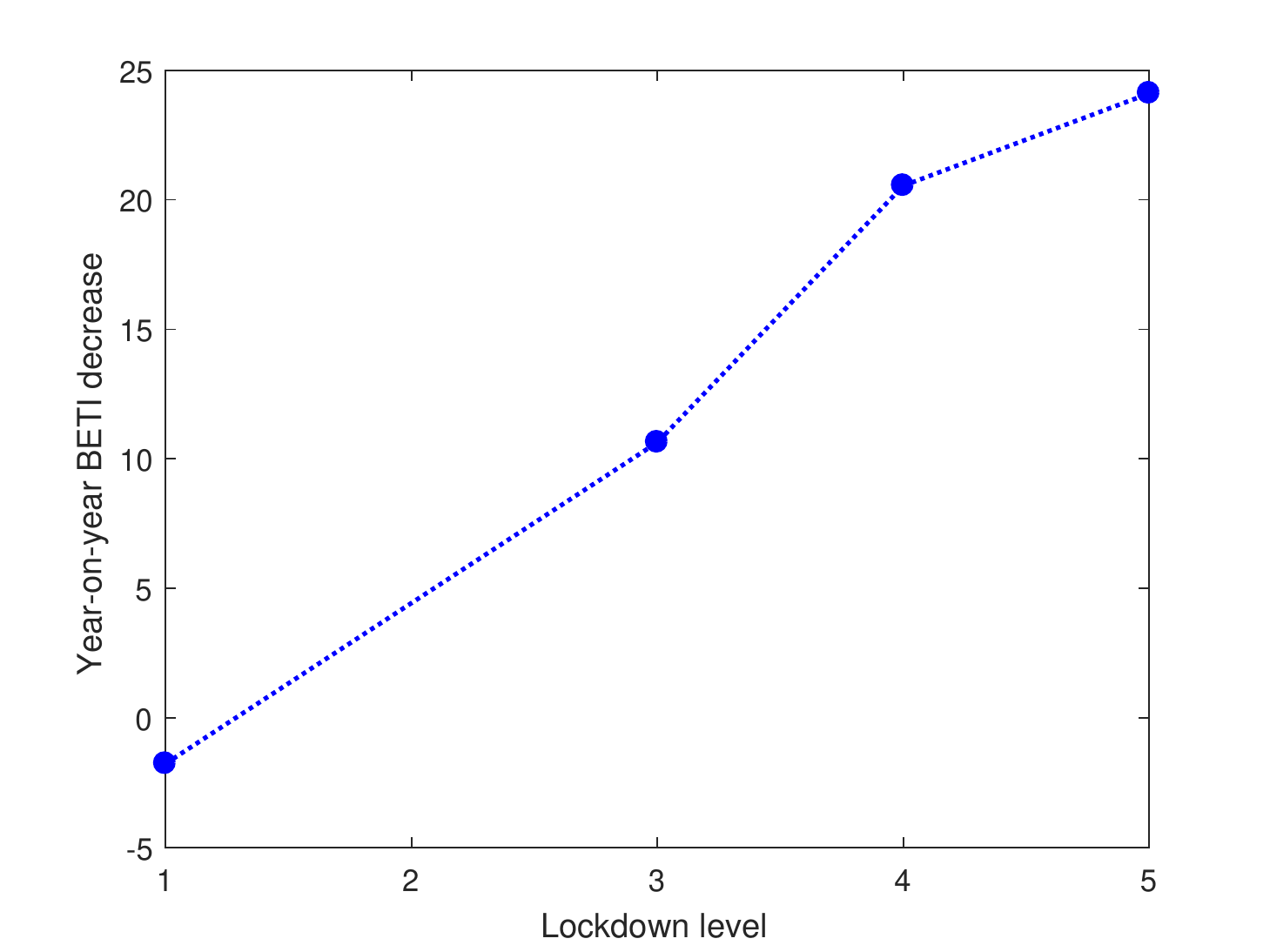}    
\caption{BETI decrease per lockdown level. Markers show where data are available and the interpolated value is not indicated with a marker. Data are taken from \cite{BankServ:20}.} 
\label{fig:economic_values_interpolated}
\end{center}
\end{figure}

There is however some seasonality connected to economic transaction volumes, and as such year-on-year values are used to gauge the impact of lockdown levels. The relative economic impact for level 1, 3, 4, and 5 are determined using the year-on-year decrease in the BETI value for March to June 2020. The value for levels 2 is interpolated from the other levels. The values and corresponding months are shown in Table~\ref{tb:economic_impact_per_lockdown_level} as well as Fig.~\ref{fig:economic_values_interpolated} to highlight the interpolation results.

\begin{table*}[t]
\begin{center}
\caption{Relative economic impact per lockdown level.}
\label{tb:economic_impact_per_lockdown_level}
\begin{tabular}{cccc}
Lockdown level & BETI decrease & Normalized economic impact & How obtained \\\hline
1 & -1.76 \% & 0.00 & BETI decrease for March 2020 \\
2 & 4.44 \%  & 0.24 & Interpolated \\
3 & 10.64 \% & 0.48 & BETI decrease for June 2020 \\
4 & 20.54 \% & 0.86 & BETI decrease for May 2020 \\
5 & 24.11 \% & 1.00 & BETI decrease for April 2020 \\
\hline
\end{tabular}
\end{center}
\end{table*}

The relative economic impact values per level are normalized to be between 0 and 1. As they are weighted in (\ref{eq:obj_function_economic_impact}) their relative values are important, and because they are included in a 2-norm calculation the absolute values need to be monotonically increasing.

This treatment of the economic impact may be somewhat simplistic, but economic data in the public domain are unfortunately limited or largely delayed. This approach does quantify the relative impact of different lockdown levels, which should be sufficient for the policy optimization.

This type of policy may reduce lockdown levels to curb economic losses, even though infection rates may be considered to be unacceptably high. This scenario is presently happening in various countries in the developing world \cite{OanhHa:20}.

\subsection{Compliance to lockdown regulations}
\label{sec:compliance}

Some residents cannot or will not endure living under continual lockdown regulations (for various reasons). \cite{BusinessTech:20} notes that 230,000 cases have been opened against South African residents for violating lockdown rules by 22 May 2020. A significant proportion of these cases were opened in the last couple of weeks before the \cite{BusinessTech:20} article was written, showing that violations become more prevalent over time. Compliance to lockdown regulations can therefore be considered as an unmeasured disturbance that impact on the parameters of the SEIQRDP model.

In this scenario (\ref{eq:obj_function_economic_impact}) is still minimized, but the compliance disturbance is introduced as a multiplicative disturbance on the expected number of contacts between people by altering the value of $\beta$; (\ref{eq:beta_per_level}) is altered to produce:

\begin{equation}
\label{eq:compliance_beta}
\beta^* (t) = \left(\beta_m (1) \cdot \left(1 - d \right) + \beta_m (l) \cdot d \right) \cdot \beta (t)
\end{equation}

where $d \in [0, 1]$ is the level of compliance to the stipulated lockdown level. When $d = 1$ the intended $\beta$-value for the associated lockdown level is obtained as in (\ref{eq:beta_per_level}). When $d = 0$ the $\beta$-value obtained is as if lockdown level 1 had been implemented.

\subsection{Control move constraint considerations}

There is a very delicate balance to be maintained when implementing policy decisions. Due regard has to be given to the effect that policies may have on the citizens of the country concerned. If not, compliance may reduce and disgruntled citizens may deliberately violate regulations or fail to keep track of which lockdown level is currently in force.

To prevent this scenario, a dynamic constraint is placed on the controller which ensures that from the present lockdown level to the end of the control horizon, the control moves must be monotonically increasing or monotonically decreasing. This constraint is enforced as:

\begin{equation} 
\label{eq:nonlcon}
\theta_c \left\{
\begin{array}{l}
 \frac{\Delta \left[ u_{k-1} , u_{k}, \ldots , u_{k+N_c-1} \right]}{\Delta t} \geq 0 \\
 \frac{\Delta \left[ u_{k-1} , u_{k}, \ldots , u_{k+N_c-1} \right]}{\Delta t} \leq 0
\end{array} 
\right. \forall ~ u_i,~ i \in [k - 1 , k + N_c - 1].
\end{equation}

This prevents excessive fluctuations of the lockdown level that the controller may otherwise seek to exploit in order to minimize the objective function. Excessive fluctuations of the lockdown levels also make it difficult to enforce and regulate the lockdown rules on a national scale as time is needed to implement any policy changes.

Once the control move (lockdown level) has been calculated, some time is needed for the nation to prepare to implement that lockdown level. Here, 1 week (7 days) of preparation time is provided. Given the blocking vector of $N_b = [7,7,7]$, this is practically achieved by fixing the first element of the control vector to the previously calculated value and then using the second element as the control move to be implemented. This is contrary to regular receding horizon control where the first element is implemented.

This implementation approach is no different from the regular receding horizon approach in the absence of model-plant mismatch and/or disturbances. No difference should therefore be expected for the initial flattening the curve and balancing lives and livelihoods policy simulations. Once the compliance disturbance is introduced however this implementation does however have some impact, given that the level of compliance when the control move is calculated may not be the same as the level of compliance when it is implemented.

\subsection{GA optimization implementation}

The controller design discussed in this section relies on a solver that can solve an objective function based on a hybrid model which contains continuous time dynamics, time varying model parameters, discontinuities in the form of discrete lockdown levels, and dynamic non-linear constraints. 

The GA is a solver that has shown good results with such mixed integer models,  non-linearities and complex non-linear constraint functions,  and convergence to a global minimum in the presence of many local minima \cite{Muller:17,Mitra:04}. It has also been successfully used as a solver for HMPC \cite{Muller:17,Botha:18}. Owing to these advantages a GA is used in this study.

The Matlab \textit{ga} function in the Global Optimization Toolbox was used to implement the GA.  The Matlab \textit{ga} function uses a set of solutions, called the population, that are calculated to minimize a fitness function. With each iteration in the GA a new generation is calculated from the old population through a mutation function (while always adhering to the upper and lower bounds as well as the inequality, equality, and non-linear constraints). The mutation function keeps the variables in the population that minimized the fitness function the most, and also defines new values according to a stochastic function. The GA is terminated when the fitness function value, fitness tolerance (i.e. the change in the fitness function value between iterations), or the maximum number of generations is exceeded \cite{Whitley:94}.

In this work the HMPC problem solves within approximately 5 seconds on a standard computing platform. As a solution is only required once a day, the execution time is not an issue. Therefore the fitness tolerance, fitness limit, population size and maximum number of generations were left as default in the Matlab \textit{ga} function. It is important to note that these parameters may be altered when the controller execution time needs to be reduced. 
The \textit{ga} function was set up using the following options and parameters:
\begin{itemize}
\item The fitness function is the objective function policy in (\ref{eq:obj_function_herd_immunity}) or (\ref{eq:obj_function_economic_impact}). 
\item The lower bound (LB) is the hard low limit for the three control moves, which is [$u_{k-1}$, 1, 1]. 
\item The upper bound (UB) is the hard high limit for the three control moves, which is [$u_{k-1}$, 5, 5]. 
\item The non-linear constraint function is implemented using (\ref{eq:nonlcon}). 
\end{itemize}

\section{Results and discussion}
\label{sec:results}

The simulation results for each scenario described in Section~\ref{sec:controller_design} is presented and discussed here. The controller design parameters and result metrics for each scenario are shown in Table~\ref{tb:control_metrics}. The metrics shown are the maximum number of active confirmed infectious cases, i.e. the maximum of Q, the number of days (from the start of the simulation on 18 April 2020) before the lockdown level is raised above 1, the number of days after the lockdown level has been raised above 1 until it is returned to 1, and the number of days in each lockdown level. The number of days in level 1 is only taken up to the point where the lockdown is finally reduced back to level 1. The number of days before lockdown is implemented is an important metric to consider as it will improve the nation's readiness for lockdown and in turn improve long term compliance.

\begin{table*}[t]
\begin{center}
\caption{Controller metrics per policy scenario.}
\label{tb:control_metrics}
\begin{tabular}{cccc}
Metric & Flattening the curve & Balancing lives and livelihoods & Compliance \\\hline \\[-7pt]
$W_s$ & $\mathrm{10^{-4}}$ & $\mathrm{10^{-4}}$ & $\mathrm{10^{-4}}$ \\
$W_u$ & $\mathrm{10^{1}}$ & $\mathrm{10^{1}}$ & $\mathrm{10^{1}}$ \\
$W_E$ & - & $\mathrm{10^{7}}$ & $\mathrm{10^{7}}$ \\\hline \\[-7pt]
Max active cases (Q) 	 & $\mathrm{1.56 \times 10^{5}}$ & $\mathrm{2.54 \times 10^{5}}$ & $\mathrm{5.76 \times 10^{5}}$ \\
Days before lockdown & 37  & 37  & 37  \\
Days in lockdown 	 & 77  & 84  & 112 \\
Days in level 5 	 & 35  & 21  & 49 \\ 
Days in level 4 	 & 7   & 7   & 21  \\
Days in level 3 	 & 7   & 28  & 28  \\
Days in level 2 	 & 28  & 28  & 14  \\
Days in level 1 	 & 37  & 37  & 37  \\[5pt]
\hline
\end{tabular}
\end{center}
\end{table*}

\subsection{Flattening the curve policy}

\begin{figure}
\begin{center}
\includegraphics[width=8.4cm]{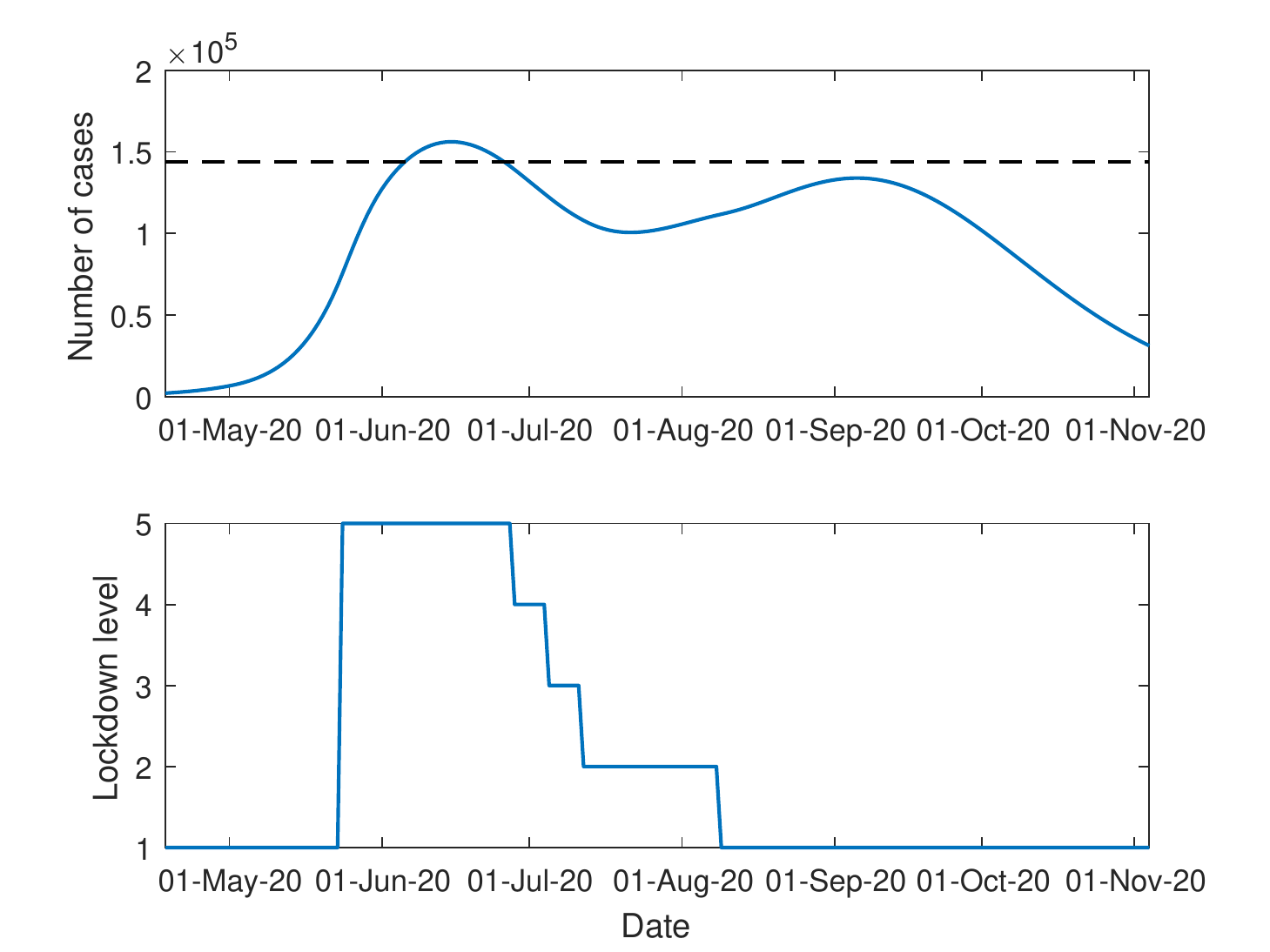}    
\caption{Simulation results for the flattening the curve policy scenario. The top panel shows the active number of confirmed infectious cases along with the number of ICU beds imposed high limit (dashed line) over time and the bottom panel shows the lockdown level over time.} 
\label{fig:Herd_immunity_control}
\end{center}
\end{figure}

The simulation result for the flattening the curve policy scenario described in Section~\ref{sec:herd_immunity} is shown in Fig.~\ref{fig:Herd_immunity_control}.

From Fig.~\ref{fig:Herd_immunity_control} it is visible that implementing lockdown at the correct time and to the correct level allows the active number of confirmed infectious cases to hardly violate the ICU capacity limit. The lockdown is however implemented relatively early (37 days after the start of the simulation on 18 April 2020) and lasts relatively long (77 days).

\subsection{Balancing lives and livelihoods policy}

\begin{figure}
\begin{center}
\includegraphics[width=8.4cm]{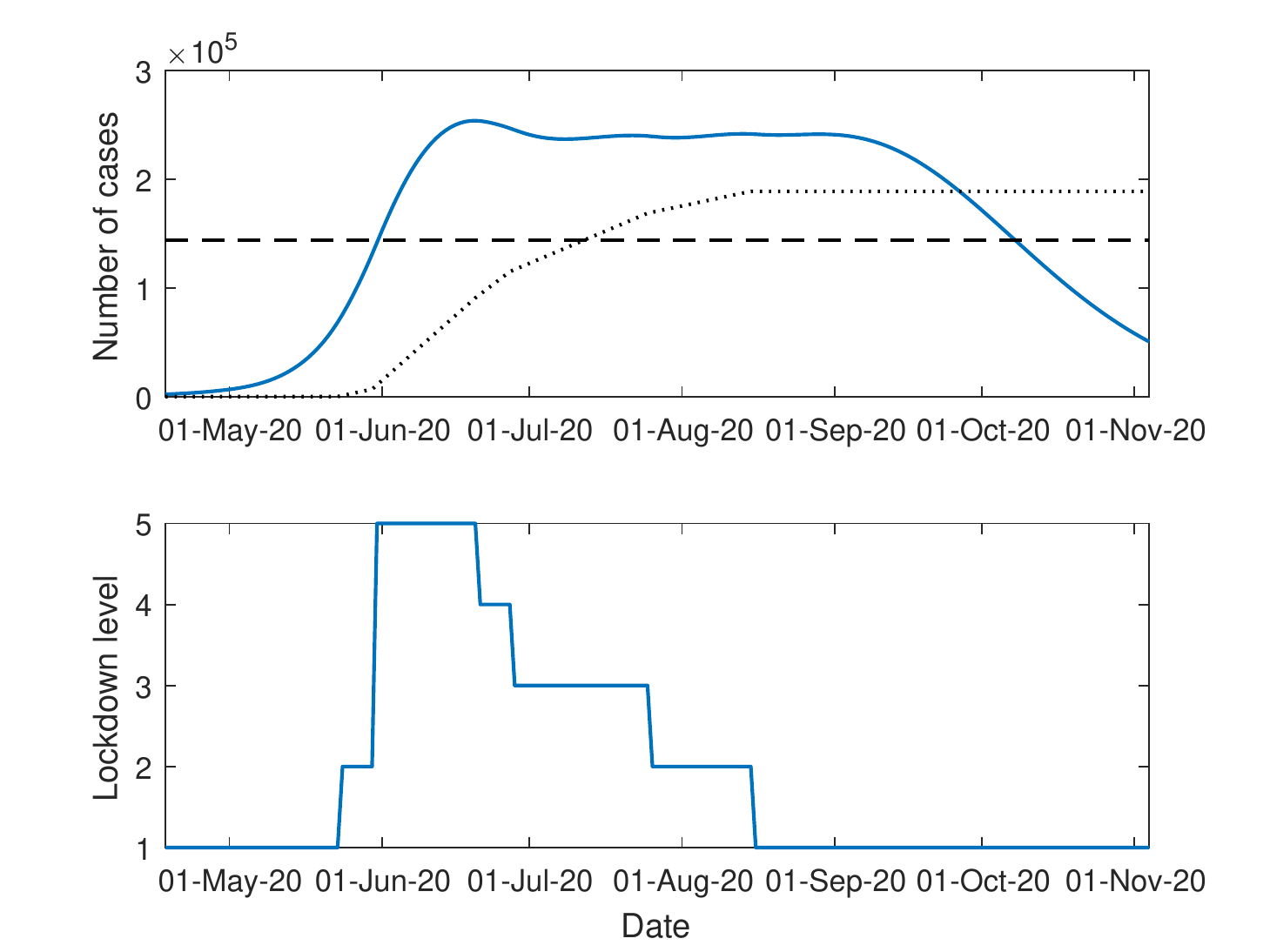}    
\caption{Simulation results for the balancing lives and livelihoods policy. The top panel shows the active number of confirmed infectious cases along with the number of ICU beds imposed high limit (dashed line) and the weighted cumulative economic impact (dotted line) over time and the bottom panel shows the lockdown level over time.} 
\label{fig:HMPC_economic_impact}
\end{center}
\end{figure}

The lockdown scenario illustrated in Fig.~\ref{fig:Herd_immunity_control} will likely have devastating effects on a South African economy that was already fragile before the onset of COVID-19. The result for the balancing lives and livelihodds policy, as described in Section~\ref{sec:economic_impact}, is shown in Fig.~\ref{fig:HMPC_economic_impact}. In this case lockdown is implemented up to level 2 and shortly after to level 5 to try and curb the spread of the virus, moving to level 5 somewhat later than in the flattening the curve scenario to alleviate some of the economic impact of lockdown. After a couple of weeks on level 5 the cumulative economic impact increases so much that the level is reduced in spite of the ICU capacity limit being exceeded. The dotted line in the top panel of Fig.~\ref{fig:HMPC_economic_impact} shows the cumulative economic impact of the lockdown that has already been implemented (as per (\ref{eq:cumulative_econ_impact})), weighted to fit onto the same scale as Q. The cumulative economic impact is already weighted ($W_E$ in (\ref{eq:cumulative_econ_impact})) meaning the original units become irrelevant. The dotted line in Fig.~\ref{fig:HMPC_economic_impact}) is therefore shown without any units, but the shape over time is informative. It is clear that as the economic impact grows, the ICU limit is allowed to be violated further in order to reduce the lockdown level and curb economic losses.

This policy balances lives and livelihoods by delaying the start of lockdown level 5 with a week and reducing the total time in level 5 by two weeks. The peak number in Fig.~\ref{fig:HMPC_economic_impact} of active confirmed infectious cases is much higher than that of Fig.~\ref{fig:Herd_immunity_control}. The advantage of this policy however is that the peak in the maximum number of cases is delayed as opposed to having no policy in place. Besides winning time, this scenario also provides an estimate of the additional number of COVID-19 specific intensive care beds required, something that healthcare authorities can use to plan their response.

This implies that if the economic impact of each level of lockdown can be quantified beforehand, temporary ICU facilities can be procured to the point where lockdown might be eased earlier to limit the cumulative economic impact while the ICU limit is still respected.

\subsection{Compliance to lockdown regulations}

\begin{figure}
\begin{center}
\includegraphics[width=8.4cm]{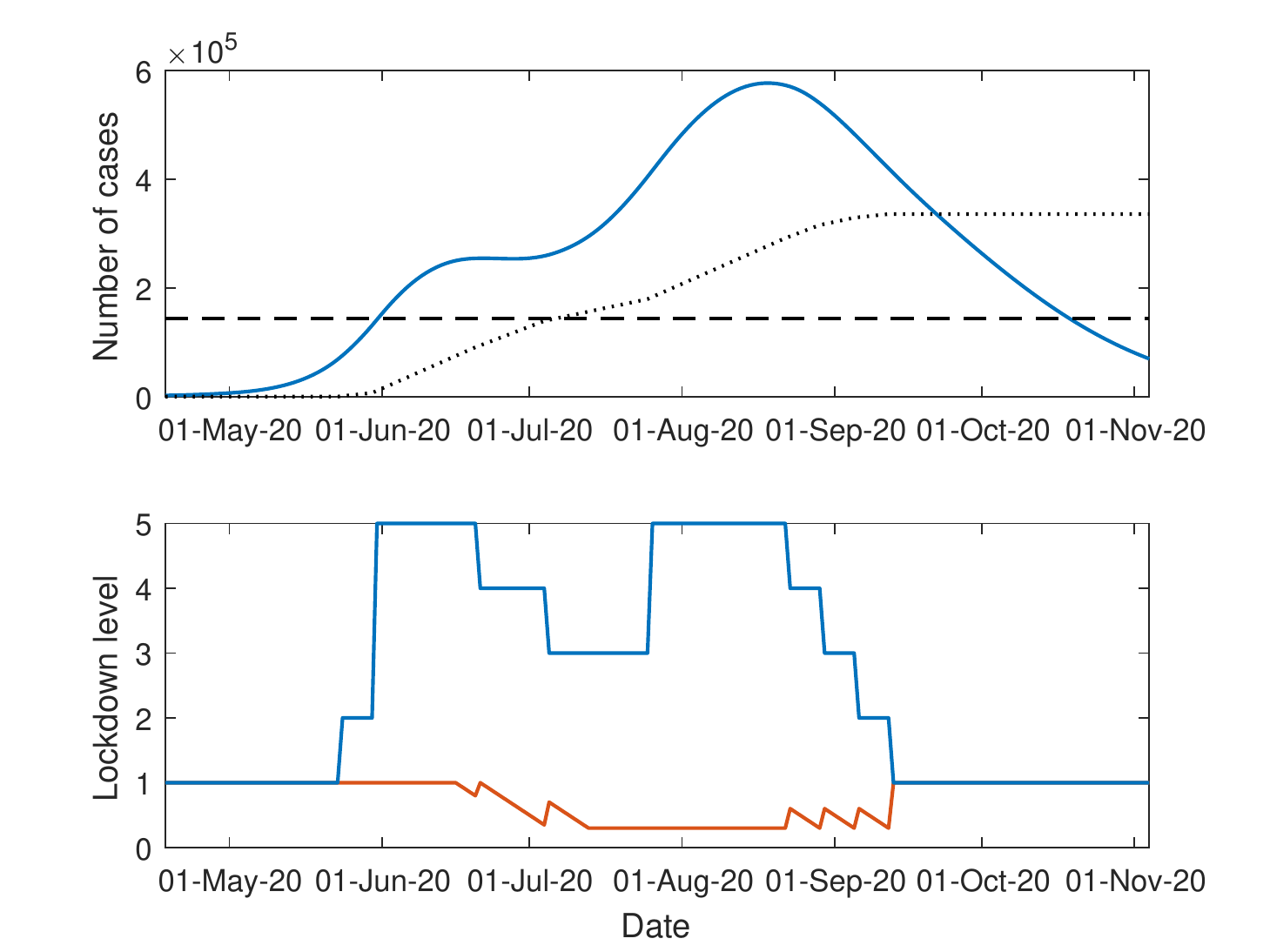}    
\caption{Simulation results for the balancing lives and livelihoods policy with waning compliance to regulations. The top panel shows the active number of confirmed infectious cases along with the number of ICU beds imposed high limit (dashed line) and weighted cumulative economic impact (dotted line) over time and the bottom panel shows the lockdown level (blue) and compliance (orange) over time.} 
\label{fig:HMPC_economic_impact_compliance}
\end{center}
\end{figure}

Given that compliance to lockdown regulations has seemingly waned over time in South Africa, the compliance parameter ($d$) in (\ref{eq:compliance_beta}) is initiated with a value of 1, and reduced linearly after a number of weeks in lockdown down to a value of 0.3 (as seen in the bottom panel of Fig.~\ref{fig:HMPC_economic_impact_compliance}). After the initial decrease the level of compliance is set to increase every time that the lockdown level is reduced. This is because there are fewer regulations on lower levels, and the level of compliance to those regulations will likely be higher initially. After the initial surge, compliance will again reduce in a linear fashion.

It is visible from Fig.~\ref{fig:HMPC_economic_impact_compliance} that the lockdown is imposed at the same time and with the same magnitude as what it was in Fig.~\ref{fig:HMPC_economic_impact}. Lockdown is also reduced from level 5 to 4 at the same time. After a couple of weeks in lockdown however the population starts to deviate from the rules and the effective number of contacts per person increases, which consequently increases the spread of the virus. To try and curb this phenomenon the controller moves the lockdown level back to 5 as it applies feedback to try and balance the ICU bed imposed limit with the economic impact of the lockdown. After another period at level 5 the cumulative economic impact has however ballooned while compliance remains relatively low. Left with large economic losses and a non-compliant population, the controller ramps down the lockdown level from 5 to 1 in a relatively short time. The compliance does increase at each reduction of the lockdown level, but this has little impact given the magnitude of the cumulative economic losses.

This policy has some resemblance to what has transpired in South Africa to date. Lockdown regulations were imposed early but relaxed while cases were still increasing owing to economic concerns. Poor compliance was noticed (see e.g. \cite{BusinessTech:20}) and after relaxing lockdown regulations some measures were re-imposed to try and curb the spread.

The preliminary version of this paper, \cite{Olivier:20b}, used a model with much higher case number predictions. This means that the lockdown durations in \cite{Olivier:20b} are much longer than those found in this work. The shapes of the control moves and outputs are however similar. Moreover, the controller hardly required any tuning changes to achieve similar results, even though the model changed significantly. This illustrates that the controller is quite robust to the model used, and that the policy concepts presented remain conceptually valid irrespective of whether the underlying model is very accurate.

\section{Limitations of the study}

Firstly, any model-based controller inherits the limitations of the model that it is based upon. At the time of modelling the spread of the virus, the SEIQRDP model appeared to be flexible enough and sufficiently representative of the underlying system to be useful. In time however it is possible that new knowledge is gained about the spread of COVID-19 that may show the SEIQRDP model to be insufficient, or that another model is superior.

Quantifying the economic impact per lockdown level may be prone to error. As stated in Section~\ref{sec:economic_impact}, it would be preferable to base the impact on GDP directly, as opposed to the proxy BETI index, but the timeliness of GDP data is limiting.

The HMPC implementation assumes full state feedback, as noted in Section~\ref{sec:controller_design}. State estimation was not implemented to limit errors that may be introduced in such a step. The states of many systems can be accurately estimated, but without a full observability analysis of the system (see e.g. \cite{LeRoux:17}) and an implementation of state estimation it is difficult to say for certain that this system lends itself to accurate estimation.

\section{Conclusion}
\label{sec:conclusion}

An epidemiological model was developed for the spread of COVID-19 in South Africa in \cite{Olivier:20}. The model was re-fitted here using data for the period from 18 April to 16 August 2020 while the country was under lockdown levels 5, 4, and 3. The model was adapted here with varying values for the spread rate ($\beta_m(l)$) under varying levels of the lockdown using more recent data. The main contribution of this paper is the implementation of the HMPC controller to determine the optimal lockdown level over time for different policy scenarios.

Under a scenario where ``flattening the curve'' is the goal, the healthcare capacity, expressed in terms of the number of available ICU beds, is largely respected, but the lockdown is severe. The detrimental cumulative economic impact of such a severe lockdown is high. A balancing lives and livelihoods policy was therefore introduced to allow for increased economic activity by reducing the lockdown level earlier. This has the effect that the ICU bed imposed limit is violated, but that more livelihoods are saved.

Lastly, compliance to lockdown regulations is added as an unmeasured disturbance. The effect of the compliance is seen through the higher number of confirmed infectious cases. To curb waning compliance the lockdown level is increased and extended to the point where the economic loss is too great and lockdown is ended rather abruptly.

The simulations therefore show how to optimally implement lockdown strategies given the main concerns and policy the implementer wants to enact.

\bibliography{ifacconf}

\begin{IEEEbiography}[{\includegraphics[width=1in,height=1.25in,clip,keepaspectratio]{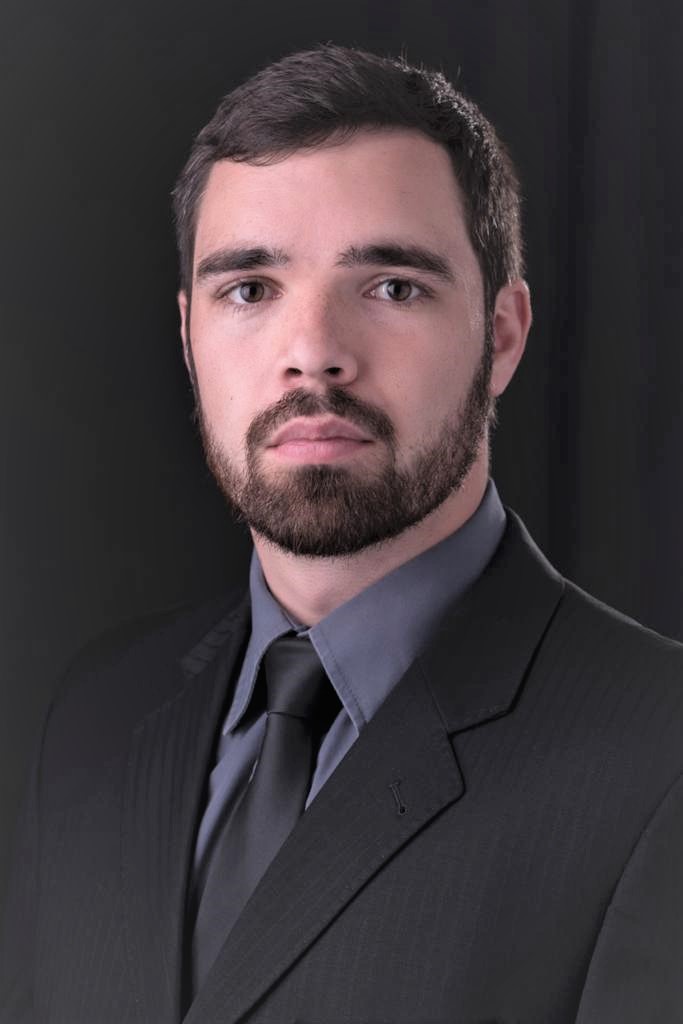}}]{Laurentz E. Olivier} was born in Pretoria, South Africa in 1987. He received B.Eng. and M.Eng. degrees in electronic engineering in 2009 and 2011 respectively, and a Ph.D. in engineering in 2016, all from the University of Pretoria (UP), South Africa.

From 2012 to 2018 he was an Advanced Process Control Engineer at the Sasol Synfuels complex in Secunda, South Africa - the largest coal to liquids factory in the world. From 2018 he has been working as a Senior Data Scientist, with a current position at Moyo Africa in Centurion, South Africa. He has also worked as a Research Associate at the Department of Electrical, Electronic, and Computer Engineering at UP from 2017. His research interests include modelling and control of minerals processing and petrochemical processes, as well as data analytics for real-world impact.

Dr. Olivier was the recipient of the Dean's prize (Gold) for the best academic achievement in Electronic Engineering in 2009, and a recipient of the University of Pretoria study-abroad bursary programme in 2011 by means of which he spent time as a visiting researcher at the University of Alberta, Canada.
\end{IEEEbiography}

\begin{IEEEbiography}[{\includegraphics[width=1in,height=1.25in,clip,keepaspectratio]{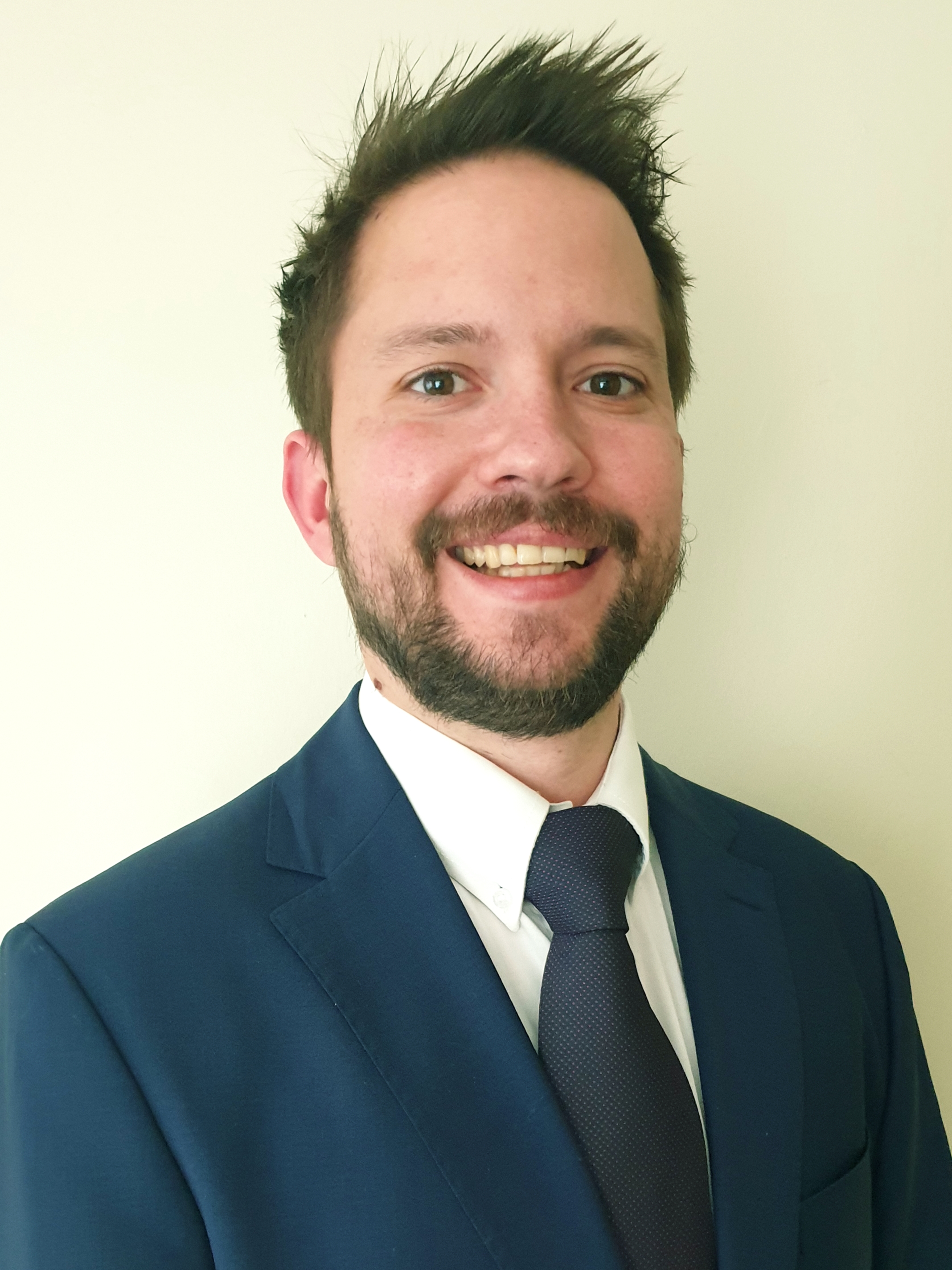}}]{Stefan Botha} was born in Durban, South Africa in 1991. He received his B.Eng. degree in 2013 and M.Eng. degree in 2018, both in electronic engineering from the University of Pretoria (UP), South Africa.

Since 2016 he has been practising as an Advanced Process Control Engineer at Sasol Synfuels in Secunda, South Africa. The Synfuels complex has an install base of more than 80 unique APC's for site-wide optimization of the petrochemical and gas value chains. His research interests include modelling and control of hybrid systems in the petrochemical and mineral processing industries, as well as advanced analytics for process fault finding and monitoring. 

Mr. Botha was the winner of the 2013 electronic engineering project competition and received the deans prize for the third best academic achiever in electronic engineering, both at the University of Pretoria. 
\end{IEEEbiography}

\begin{IEEEbiography}[{\includegraphics[width=1in,height=1.25in,clip,keepaspectratio]{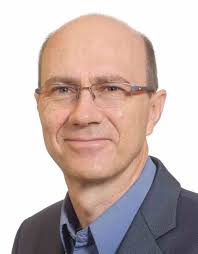}}]{Ian K. Craig} (M'86--SM'97) received the B.Eng. degree in electronic engineering from the University of
Pretoria (UP), Pretoria, South Africa, the S.M. degree from the Massachusetts Institute of Technology,
Cambridge, and the Ph.D. and M.B.A. degrees from
the University of the Witwatersrand, Johannesburg,
South Africa.

He has been Professor and Section Head: Control
Systems in the Department of Electrical, Electronic
and Computer Engineering at UP since 1995. Previously he was Group Leader in the Measurement Control Division of Mintek, where he was involved in the design and implementation of advanced controllers for the mineral processing industry. His research
interests include industrial process control as applied to metals and mineral processing, the economic evaluation of control systems, and lately the control of
biomedical systems.

Prof. Craig was Editor-in-Chief of the journal Control Engineering Practice from 2005-2010 and President of the International Federation of Automatic Control (IFAC) from 2011-2014. He is an IFAC Advisor, a fellow of the South African Academy of Engineering, and a registered Professional Engineer in South Africa.
\end{IEEEbiography}

\EOD

\end{document}